\documentclass[a4paper]{article}

\usepackage{LaTeX/INTERSPEECH2022}
\usepackage{tipa}
\usepackage[xcdraw]{xcolor}

\title{Nonwords Pronunciation Classification in Language Development Tests \\for Preschool Children}
\name{Ilja Baumann$^1$, Dominik Wagner$^1$, Sebastian Bayerl$^1$, Tobias Bocklet$^{1,2}$}

\address{
  $^1$Technische Hochschule Nürnberg Georg Simon Ohm, Germany\\
  $^2$Intel Labs
  }
\email{\{ilja.baumann, dominik.wagner, sebastian.bayerl, tobias.bocklet\}@th-nuernberg.de} 

\begin{document}
\maketitle


\begin{abstract}
This work aims to automatically evaluate whether the language development of children is age-appropriate.
Validated speech and language tests are used for this purpose to test the auditory memory. In this work, the task is to determine whether spoken nonwords have been uttered correctly.
We compare different approaches that are motivated to model specific language structures:
Low-level features (FFT), speaker embeddings (ECAPA-TDNN), grapheme-motivated embeddings (wav2vec 2.0), and phonetic embeddings in form of senones (ASR acoustic model).
Each of the approaches provides input for VGG-like 5-layer CNN classifiers.
We also examine the adaptation per nonword.
The evaluation of the proposed systems was performed using recordings from different kindergartens of spoken nonwords.
ECAPA-TDNN and low-level FFT features do not explicitly model phonetic information; wav2vec2.0 is trained on grapheme labels, our ASR acoustic model features contain (sub-)phonetic information. 
We found that the more granular the phonetic modeling is, the higher are the achieved recognition rates. 
The best system trained on ASR acoustic model features with VTLN achieved an accuracy of 89.4\% and an area under the ROC (Receiver Operating Characteristic) curve (AUC) of 0.923.
This corresponds to an improvement in accuracy of 20.2\% and AUC of 0.309 relative compared to the FFT-baseline.


\end{abstract}

\noindent\textbf{Index Terms}: children's speech, speech assessment, transfer learning

\section{Introduction}
Language competence is one of the key skills for later academic success. A need for support should be identified as soon as possible \cite{norbury2016younger}. For this reason, language assessment as part of a school enrollment examination is mandatory in Germany. Professional speech therapists perform the assessment. 
One of the aspects to be tested is auditory memory, with phonological representability and short-term memory span being of particular importance.
In German-speaking countries, among others, the SETK (Sprachentwicklungstest für Kinder – Language Development Test for Children) \cite{grimm2001setk} test is used to diagnose language processing abilities and auditory memory capacity. The test is designed for children in the critical age range 3;0 – 5;11 years or older children with known developmental disabilities. The SETK is divided into subtests, each of which can be assigned to language comprehension, language production or language memory. In order to determine the level of language attained and to relate it causally to auditory memory performance, phonological working memory for nonwords or made-up words is a crucial factor \cite{baddeley2003working}.
The fact that children with speech disorders repeat nonwords less accurately than children of the same age with normal speech development was proven by various studies \cite{ gathercole1990phonological, montgomery1995sentence}.
The phonological working memory subtest for nonwords as part of the SETK test is evaluated in this paper. The aim is to classify children's recited nonwords to determine whether the pronunciation was correct or incorrect. Different systems for automatic classification are compared and evaluated to support speech therapists during a test in a screening setup. Beginning with modeling of raw features, utterance embeddings, grapheme-motivated embeddings, and finally phonetic embeddings.

Convolutional Neural Networks (CNNs) proved to be very effective in modeling aspects of speech and have been successfully applied to downstream tasks, such as speech emotion recognition \cite{huang2014speech}.
In \cite{hershey2017cnn}, architectures such as AlexNet, VGG, Inception, and ResNet from the image domain were primarily used to classify audio. Results indicate that neural embeddings extracted from these models perform better than raw feature inputs.
In \cite{el2019transfer}, VGGish, a pre-trained deep CNN was used as a feature extractor to train a classifier based on the AudioSet dataset \cite{gemmeke2017audio}. The extracted features are vastly superior over MFCCs with 30 times fewer features to represent the data.
Phoneme pronunciation classification was investigated in \cite{asif2021approach}, using CNN architectures and fine-tuning approaches.
In \cite{mohammed2020voice}, voice pathology detection is performed using CNN models.
X-Vectors \cite{snyder2018x} or ECAPA-TDNN as enhanced architecture, were originally introduced for speaker verification. The architectures have been successfully used in the areas of speaker diarization \cite{dawalatabad2021ecapa}, dialect classification \cite{kethireddy2022deep}, pathologic speech assessment \cite{Pappagari20-USO, Scheurer21-AXO} and other paralinguistic tasks, e.g., emotion recognition \cite{Pappagari20-XME}.
Wav2vec 2.0 \cite{baevski2020wav2vec} models have been successfully applied to downstream speech processing tasks such as emotion recognition \cite{pepino2021emotion} or detection of Alzheimer's speech \cite{Balagopalan21-CAA}.

We use a VGG-like CNN system as a backend to classify the pronunciation of spoken nonwords.
We also investigate fine-tuning for word-optimized classification.
As input, we use mel spectrograms and outputs from pretrained state-of-the-art architectures in related domains. These are motivated to extract and model different levels of speech structure. We in particular use ECAPA-TDNN \cite{desplanques2020ecapa}, wav2vec 2.0 \cite{baevski2020wav2vec} and a Kaldi-based \cite{povey2011kaldi} TDNN acoustic model for ASR.
Whole utterance embeddings are extracted with a pre-trained ECAPA-TDNN for speaker verification. The embeddings represent a holistic expression without explicit modeling of phonetic information.
Grapheme level embeddings are extracted via wav2vec 2.0; embeddings after each transformer block are compared.
Finally, we use a phonetically motivated embedding generator in form of an ASR acoustic model, i.e., the senone outputs from a Kaldi-based time-delay neural network (TDNN) acoustic model. The embeddings of this modeling contain (sub-)phonetic information.



\section{Data}
The SETK subtest for phonological working memory used in this work consists of nonwords that have to be repeated by the children.
Performance is evaluated based on the number of correct and incorrect repetitions. 
The ability to capture unfamiliar phonetic patterns, store them short-term in working memory, and recall them immediately is tested.
Phonological awareness is the underlying process that describes children's ability to identify sublexical units based on holistic representations. 
In addition, the repetition of nonwords tests the phonological short-term memory since phonological patterns which are not known have no entry in the lexical memory \cite{metsala1999young}.
Seven nonwords must be repeated during the test, which differ in complexity, length, and number of syllables. 
The test begins with 2-syllable nonwords and gradually progresses to 5 syllables. The nonwords are: \textit{Maluk \textipa{[ma\textlengthmark-l\textupsilon k]}}, \textit{Bilop \textipa{[bi\textlengthmark-l\textopeno p]}}, \textit{Ronterklabe \textipa{[\textinvscr\textopeno n-t\textturna-kla\textlengthmark-b\textschwa]}}, \textit{Glösterkeit \textipa{[gl\o\textlengthmark s-t\textturna-ka\textsci t]}}, \textit{Seregropist \textipa{[ze\textlengthmark-\textinvscr e-g\textinvscr o\textlengthmark-p\textsci st]}}, \textit{Pristobierichkeit \textipa{[p\textinvscr\textsci s-to\textlengthmark-bi\textlengthmark-\textinvscr\textsci \c{c}-ka\textsci t]}} and \textit{Kabusaniker \textipa{[ka-bu\textlengthmark-za-ni\textlengthmark-k\textturna]}}.

Audio recordings were collected by the authors of \cite{bocklet2009automatic} in various German kindergartens. The recordings were obtained by experienced speech therapists in the respective kindergartens on site. Each child was tested and recorded individually. For each nonword, a recording of the spoken nonword by a speech therapist was played to provide the same conditions for each recording. 
The final dataset consists of 984 utterances or nonwords from 140 children. 
The data was labeled by two expert raters. 
Each nonword was rated as correct or incorrect, with an additional confidence rating, confident or uncertain, since some statements were very unclear. 
Only utterances for which both raters were confident that they labeled correctly and agreed on the rating were selected, resulting in 651 utterances from 139 children. 
We divide the data into a training and test set (random 75\% / 25\% split). 25\% of the test set is taken for validation.
We removed silence, other voices, and noise from the recordings using rVAD \cite{tan2020rvad} before and after each nonword is repeated.

\section{Modeling Techniques}

\subsection{Classification system}
\label{sec:classification-system}
\begin{figure}[t]
  \centering
  \includegraphics[width=\linewidth]{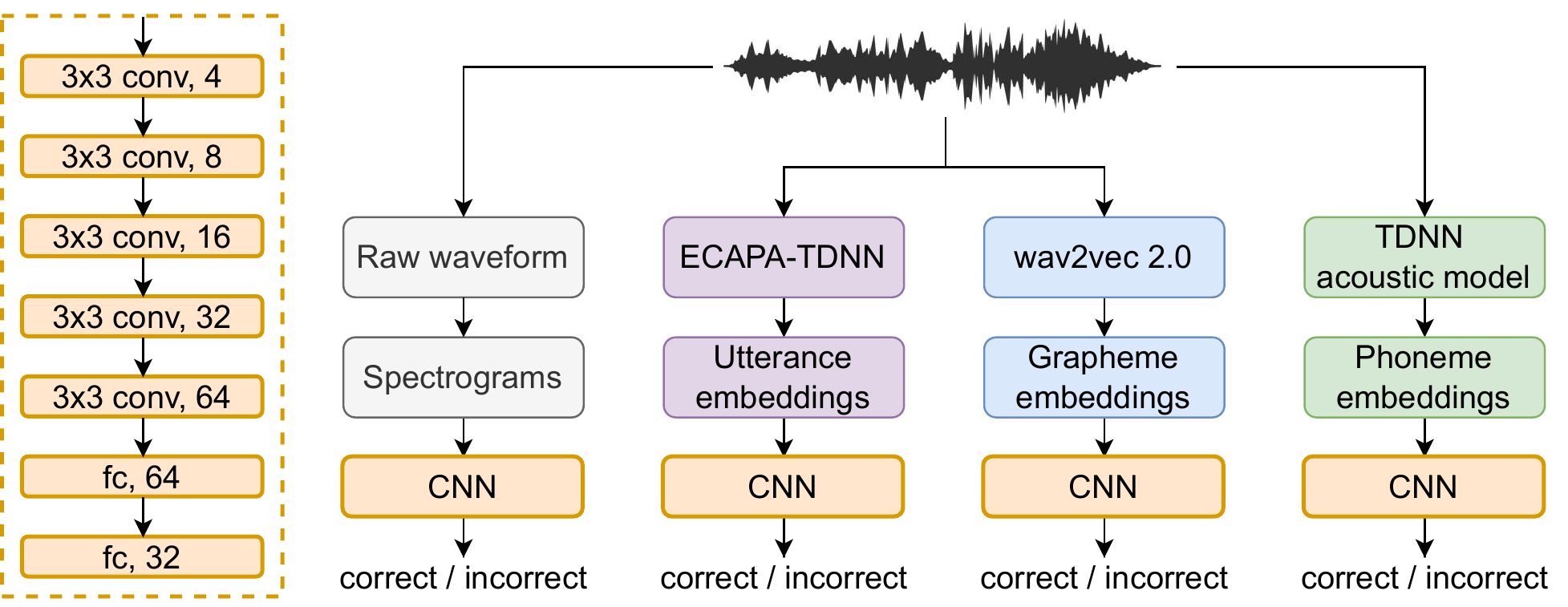}
  \caption{The proposed systems use identical CNN architectures to classify the nonwords. Features are extracted more and more granular according to the speech structure. }
  \label{fig:architectures}
  \vspace{-3mm}
\end{figure}

Our baseline is a CNN system trained on spectrograms, inspired by the VGG \cite{simonyan2014very} network used for image classification. The network consists of a series of convolution and activation layers, followed by batch normalization and a max-pooling layer.
Similar to VGG, the convolutional layers have small filters with a kernel size of 3×3 and a stride of 1. ReLU was used as activation after convolution. 
Our best-performing base model had 5 layers and 4 filter kernels in the first layer; filter kernels are doubled layer by layer. The left-hand side of Figure \ref{fig:architectures} shows the CNN architecture. Convolutional layers are followed by fully connected layers containing the number of channels corresponding to the number of filter kernels of the last convolutional layer.

The proposed network structure was used to train with different features in each case, extracted with ECAPA-TDNN, wav2vec 2.0, and Kaldi TDNN acoustic models.
Figure \ref{fig:architectures} illustrates, beside the CNN classifier architecture, the proposed classifiers trained on different features, modeling specific aspects of speech.
Only the input layer was modified for the corresponding feature dimensions.

In addition, different fine-tuning methods were explored in the baseline setup.
The fine-tuning was performed on the respective word-independent models trained on all words. 
We trained word-dependent models in 3 different ways:
\\
1) The last linear layer was retrained for each nonword, the remaining layers were frozen. \\
2) The whole pre-trained model was fine-tuned with a lower learning rate of $10^{-5}$. \\
3) An additional linear layer with 16 nodes was appended to the model, all other layers of the pre-trained model were frozen.\\
For comparison, we trained a word-dependent model for each nonword from scratch, i.e., without pre-training.

\subsection{Features}
\subsubsection{Low-level features} 
We first used low-level features for the training of the baseline system.
We converted the input audio waveform of the whole utterance into a sequence of 128-dimensional mel spectrograms.
The mel spectrograms were obtained through a short-time Fourier transform (STFT) using a 25ms frame size, 10ms frame hop, and a Hann window function.
The resulting sequences were zero-padded to the same length according to the longest element. 
By using the spectrograms, we attempt to model each nonword as a holistic representation that does not contain any underlying information of the speech.

\subsubsection{Utterance embeddings} 
ECAPA-TDNN \cite{desplanques2020ecapa} is an architecture based on the x-vectors \cite{snyder2018x} topology, a Time Delay Neural Network (TDNN) that applies statistics pooling to project variable-length utterances into fixed-length speaker embeddings. Several enhancements were introduced to create more robust speaker embeddings.
A channel- and context-dependent attention mechanism performs pooling, allowing the network to attend to different frames per channel. In addition, global context information is added to local operating convolutional blocks by using 1-dimensional squeeze-excitation (SE) blocks that rescale the channels of intermediate frame-level feature maps. 
1-dimensional Res2Net-blocks improve performance and reduce the total number of parameters. 
Features are concatenated with intermediate outputs from previous layers before the attentive statistics pooling layer.

The ECAPA-TDNN \cite{ravanelli2021speechbrain} was trained on the development part of the VoxCeleb2 dataset with 5994 speakers. 80-dimensional MFCCs have been generated as input features. Different augmentations were applied to the training data, including time-domain SpecAugment, reverb, and noise. 
Using ECAPA-TDNN we extracted 192-dimensional embeddings over the entire utterances. 
We trained a simple classifier with two linear layers and a 256-dimensionality of output space.
The whole pre-trained classifier was fine-tuned for each nonword.
Similar to spectrograms, we use these embeddings to model a holistic representation of the words, as ECAPA-TDNN was trained on speaker labels without phonetic information.

\subsubsection{Grapheme embeddings} 
Wav2vec 2.0 uses self-supervised learning and subsequent fine-tuning to make speech recognition more robust using less labeled data. 
The raw waveform is normalized and fed into a feature encoder, consisting of several blocks of temporal convolutions, layer normalization, and gaussian error linear unit (GELU) activation function.
The output of the feature encoder is fed into a transformer network.
A convolutional layer is used to derive the relative position instead of positional embeddings.
Using product quantization, the output of the feature encoder is also discretized to a finite set of speech representations.
Since a fraction of time steps in the latent feature encoder space is masked, a contrastive task is given.
The model has to identify the true quantized representation for a masked time step. 
Fine-tuning of pre-trained models for speech recognition is achieved by adding a randomly initialized linear projection on top of the context network. 

We used a wav2vec 2.0 model pre-trained on unlabeled audio from the LibriSpeech dataset \cite{panayotov2015librispeech} and fine-tuned for ASR on the same audio with the corresponding transcripts.
We extracted 768-dimensional features for every 20ms of audio. 
Wav2vec 2.0 is trained on unlabeled data and fine-tuned with an additional layer for ASR to predict characters. 
This results in embeddings of nonwords at the grapheme-motivated level of speech.

\subsubsection{Phonetic embeddings} 
We finally use the ASR recipe Kaldi Tuda-De provided in \cite{milde2018open} available for German to extract phonetic embeddings. The model is trained on freely available data resources, including the Spoken Wikipedia Corpora (SWC), the German subset of m-ailabs, the German Commonvoice Corpus, and Tuda-De. The final result is about 1050 hours of training data.
We trained a hybrid DNN-HMM model for ASR and used a TDNN as an acoustic model following the chain recipe in \cite{milde2018open}. 
Since our goal is to classify the nonwords, we do not use the whole ASR system but only the outputs of the acoustic model to train a classifier based on these. The input to the TDNN acoustic model is composed of two types of features: 40-dimensional MFCCs extracted from a 25ms window and 10ms frameshift and 100-dimensional i-vectors from chunks of 150 consecutive frames. The final input is a 220-dimensional concatenated vector consisting of three consecutive MFCC vectors and i-vector. For decorrelation, LDA is applied on the features without changing the dimensionality.
The TDNN stack consists of 6 blocks that operate on the 2048 dimensional output of the previous block. Each TDNN block is a succession of 1-dimensional temporal convolutions except in the first block and a consecutive process of ReLU activation and batch normalization. The contexts differ from block to block and use sub-sampling in the last 3 blocks, skipping temporal frames to access a broader context. The network is trained with two outputs for better performance, a chain-based LF-MMI \cite{povey2016purely} criterion and cross-entropy criterion. However, only the chain-based output without softmax is used for inference, since the model is trained with sequence objective. The output of the acoustic model is 3510-dimensional and consists of posterior probabilities of the acoustic states.
Based on the extracted senones we trained a word-independent classifier and fine-tuned it for word-dependent models.

Due to the shorter vocal tract and smaller vocal folds, children's utterances have a higher fundamental frequency than adults \cite{fitch1999vocaltract}.
To normalize the spectral distribution of different speakers, vocal tract length normalization (VTLN) was used as a feature-level transformation.
In this process, the positions of the center frequencies of the triangular frequency bins are moved.
We applied VTLN to match the fundamental frequency of the input to the training data as closely as possible. Using FCN-f0 \cite{ardaillon2019fcnf0} we calculated the average fundamental frequency on the original training data of the Kaldi ASR model. Afterward, we calculated a warp factor per speaker to obtain more robust scores based on the short utterances.
We finally applied the warp factor to each speaker's utterances before extracting the senones.
The resulting features represent the most granular unit of speech with the phonetic embeddings in this work.

\section{Experiments and Results}
The classifiers were trained with TensorFlow \cite{abadi2016tensorflow} using the Adam \cite{kingma2014adam} optimizer. Batch normalization \cite{ioffe2015batch} was applied after each convolutional layer. In the last linear layers, dropout was used with a probability of 0.5. A sigmoid layer was used as the final layer in all models, binary cross-entropy as the loss function. Class weights were applied according to the respective distribution of labels in training to compensate data imbalance, especially in word-specific models.
The test set was constant across all experiments.
We report accuracy, precision, recall, and AUC, the area under the Receiver Operating Characteristic (ROC) curve \cite{fawcett2004roc} across the 7 nonwords. 

We first trained word-independent classifiers on the different features and fine-tuned these further.
The SETK test is established as a screening test, and one nonword is tested after the other by a speech therapist. 
Due to this, it is always known which nonword is being tested, and therefore word-dependent classifiers can be used.
We investigated different methods to implement word-dependent systems in \ref{sec:classification-system}. 
Thus, for each type of feature, one word-independent system and 7 word-dependent systems were created.

\subsection{Results}
\subsubsection{Low-level features}
We trained the baseline classifier on spectrograms with a learning rate of $10^{-3}$ for 100 epochs and early stopping with patience of 10 epochs. The entire training set was used as input, which provides the word-independent model.
The results are listed in Table \ref{tab:baseline-results}, including the proposed fine-tuning methods for word-dependent systems. Word-dependent models performed best, with fine-tuning of the word-independent model. Training from scratch for each nonword provided similar performance. Only the models fine-tuned on all layers provided an AUC over 0.7. Nevertheless, the precision of the best models remains low, with 0.693 across all nonwords. 
The best approach, fine-tuning all layers of a pre-trained word-independent model, was used for further experiments with different embedding types of speech.

\begin{table}[th]
  \caption{Baseline CNN systems trained on mel spectrograms.}
  \label{tab:baseline-results}
  \centering
  \begin{tabular}{lllll}
    \toprule
    \textbf{System}     & Accuracy & Precision & Recall & AUC \\
    \midrule
    Word-independent    & 0.741 & 0.627 & 0.722 & 0.640 \\\cmidrule{1-5}
    \multicolumn{5}{l}{Word-dependent} \\
    - FT last layer     & 0.682 & 0.660 & 0.843 & 0.525 \\
    - FT all layers     & 0.744 & 0.693 & 0.737 & 0.705 \\
    - FT add. layer     & 0.704 & 0.611 & 0.586 & 0.554 \\
    From scratch      & 0.720 & 0.669 & 0.758 & 0.689 \\
    \bottomrule
  \end{tabular}
  \vspace{-3mm}
\end{table}

\subsubsection{Utterance embeddings} 
We used ECAPA-TDNN extracted embeddings to train a word-independent model and fine-tuned it.
Results in Table \ref{tab:ecapa-results} show that the fine-tuned models are similar to the baseline in accuracy, precision, and recall. However, low AUC indicates that due to the short utterances, not enough meaningful information could be extracted using the pre-trained ECAPA-TDNN. An examination of the individual nonword classes revealed that AUC was particularly low for short words. It is noticeable for \textit{Bilop} with an AUC of 0.523, whereas \textit{Pristobierichkeit} and \textit{Kabusaniker} result in 0.758 and 0.803.

\begin{table}[th]
  \caption{ECAPA-TDNN word-independent and fine-tuned word-independent systems.}
  \label{tab:ecapa-results}
  \centering
  \begin{tabular}{lllll}
    \toprule
    \textbf{System}       & Accuracy & Precision & Recall & AUC \\
    \midrule
    
    Word-independent      & 0.610 & 0.630 & 0.632 & 0.505 \\
    Word-dependent        & 0.727 & 0.746 & 0.790 & 0.689 \\
    \bottomrule
  \end{tabular}
\end{table}

\subsubsection{Grapheme embeddings}
\begin{figure}[t]
  \centering
  \includegraphics[width=0.9\linewidth]{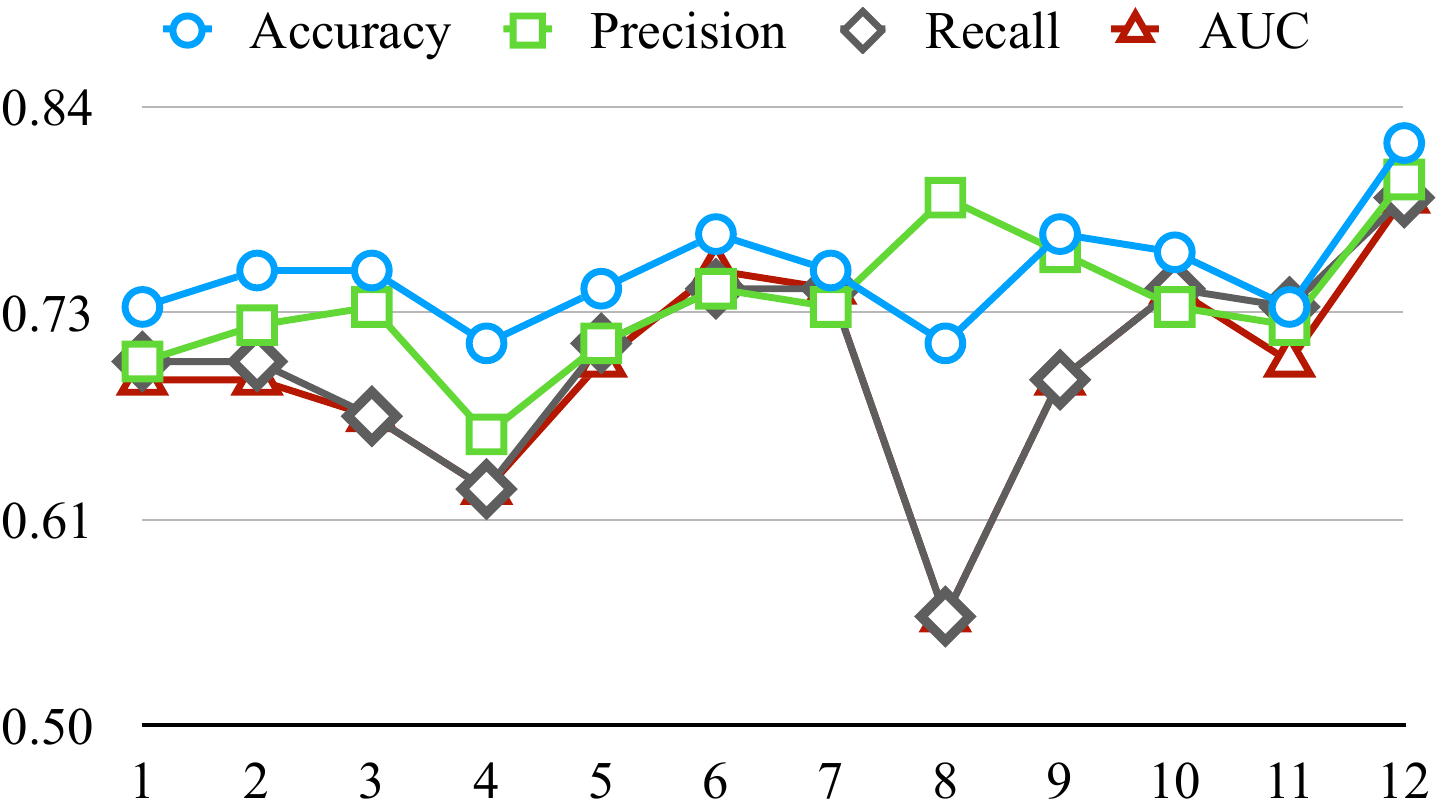}
  \caption{Wav2vec 2.0 transformer blocks 1-12 with the respective weighted averaged metrics for word-independent models.}
  \label{fig:w2v-layers}
  \vspace{-3mm}
\end{figure}

Using grapheme-motivated embeddings from wav2vec 2.0, we trained one word-independent model for each output of the 12 transformer blocks to find the most relevant for the classification of nonwords.
Figure \ref{fig:w2v-layers} shows the result for each transformer block output features for word-independent models.
We got the best results from the last transformer block. 
The higher the processing hierarchy of the respective transformer blocks, the better were the results.
This is consistent with the processing hierarchy, where the higher-level modeling is more relevant to phonetic structures. 
We used the wav2vec 2.0 outputs of the last transformer block to pre-train the word-independent model, on which the word-specific fine-tuning was applied.
Table \ref{tab:wav2vec-results} shows the results of the wav2vec 2.0 feature models. By fine-tuning to word-dependent classifiers, a precision, recall, and AUC around 0.8 is achieved. The lowest AUC for the fine-tuned models is 0.683 for one of the longest nonwords \textit{Seregropist}. The variability of the performance depending on the nonword class is smaller compared to the baseline system. 

\begin{table}[th]
  \caption{Wav2vec 2.0 word-independent and fine-tuned word-independent systems.}
  \label{tab:wav2vec-results}
  \centering
  \begin{tabular}{lllll}
    \toprule
    \textbf{System}       & Accuracy & Precision & Recall & AUC \\
    \midrule
    Word-independent        & 0.766 & 0.607 & 0.633 & 0.707 \\
    Word-dependent          & 0.827 & 0.800 & 0.794 & 0.794 \\
    \bottomrule
  \end{tabular}
  \vspace{-3mm}
\end{table}

\subsubsection{Phonetic embeddings}
We used the senones of the Kaldi-based acoustic model to train two word-independent models, with and without VTLN applied, and fine-tuned these.
Results in Table \ref{tab:kaldi-results} show that the word-independent model outperforms the other modeling techniques considered in this work. Fine-tuning leads to further improvements.
Between the word-independent models with and without VTLN, the main improvement was in AUC from 0.783 to 0.848.
Especially in fine-tuning, the use of VTLN showed a major impact.
Also, more complex and long nonwords like \textit{Pristobierichkeit} yield solid results in precision, recall, and AUC with 0.8, 1.0, 0.96.
Compared to the word-dependent models with wav2vec 2.0 features, phonetic embeddings lead to improvements in precision, recall, and AUC from 6-18\% relative.

\begin{table}[th]
  \caption{Word-independent and fine-tuned word-independent systems based on TDNN ASR acoustic model features.}
  \label{tab:kaldi-results}
  \centering
  \begin{tabular}{lllll}
    \toprule
    \textbf{System}          & Accuracy & Precision & Recall & AUC \\
    \midrule
    Word-independent         & 0.815 & 0.849 & 0.705 & 0.783 \\
    Word-dependent           & 0.842 & 0.863 & 0.835 & 0.848 \\\cmidrule{1-5}
    \multicolumn{5}{l}{VTLN} \\
    Word-independent         & 0.814 & 0.854 & 0.727 & 0.848 \\
    Word-dependent       & 0.894 & 0.849 & 0.934 & 0.923 \\
    \bottomrule
  \end{tabular}
  \vspace{-3mm}
\end{table}

\subsection{Discussion}
Our findings show that the classification accuracy increases gradually with the phonetic modeling capabilities of the different systems.
At the same time, the variability in the evaluation of individual nonwords decreases.
The more detailed phonetic representations are modeled, the more meaningful the features and the more robust the recognizers are. For the task of evaluating specific nonwords, a precise modeling of phonetic information is more important than for speaker-holistic tasks, e.g., accent, emotion, or presence/absence of certain diseases.
Spectrograms and ECAPA-TDNN embeddings, modeling the holistic statement, have provided the lowest results on our task with the highest performance variability across the different nonwords (AUC standard deviation (SD) of 0.218 and 0.132). 
The use of grapheme-motivated features (wav2vec 2.0) results in improvements of all metrics, especially AUC.
The variability of individual nonwords performance is lower compared to ECAPA-TDNN embeddings and spectrograms (AUC SD of 0.114).
This is because more detailed structures of speech are modeled over time.
Senones as (sub-)phonetic level features from the ASR acoustic model performed best. Probably also because more significant information can be obtained from an even more granular modeling of speech compared to wav2vec 2.0 features. The metrics variability across nonwords is also the lowest (AUC SD of 0.050).
The more detailed the phonetic information seems to be modeled, the better the results.

\section{Conclusion}
We explored different ways of extracting and modeling features for nonword classification. 
Low-level features, speaker embeddings, grapheme-motivated embeddings, and phonetically-motivated embeddings.
We proposed a 5 layer VGG-like CNN as the classifier for the different feature levels of the extracted speech.
We evaluated the models on nonwords spoken by children and showed that wav2vec 2.0 and especially TDNN acoustic models for ASR provide robust features for nonword classification.
TDNN acoustic model with VTLN at feature-level outperformed other embeddings while using an identical CNN classifier system. The best model trained with phonetic embeddings achieved an accuracy of 89.4\%, precision of 0.849, recall of 0.934, and AUC of 0.923.
We have shown the progress of performance over the different levels of speech modeling and how it increases the more granular the speech structure is.
To the best of our knowledge, this is the first comparison of modeling different detail levels of speech and using senones for classification.


\section{Acknowledgements}
We thank Birgit Müller-Kolmstetter for labeling and the support from the perspective of a speech therapist.

\newpage

\bibliographystyle{LaTeX/IEEEtran}

\bibliography{LaTeX/mybib}

\end{document}